# HADES: a High Amperage Driver for Extreme States

P.-A. Gourdain, M. Evans, B. Foy, D. Mager

*Extreme State Physics Laboratory, Department of Physics and Astronomy, University of Rochester, Rochester, NY 14618*

R. McBride

*Nuclear Engineering and Radiological Sciences, University of Michigan, Ann Arbor, Ann Arbor, MI 48109*

R. Spielman

*Department of Physics, Nuclear and Electrical Engineering, Idaho State University, Pocatello, ID 83209*

**Abstract:**

Linear transformer drivers (LTD) allow to greatly reduce the size of pulsed-power drivers while increasing their efficiency and repetition rate. However, limitations on the capacitors voltage and current exist, mostly driven by technological imperatives. As a result, LTD required to be connected in series *and* in parallel to form a practical pulsed-power generator. The High Amperage Driver for Extreme States (HADES) proposed here can generate 250 GW of power, i.e. a 1 MA on current with a rise time of 150 ns into a 20nH load. The key feature of the driver is its size. The machine footprint is less than 5m$^2$ and its height is less than 1.6m. This reduction in footprint, compared to other 1 MA LTD based-systems, comes from a compact transmission line geometry which allow to connect LTDs is series and in parallel efficiently. This paper summarizes this design and describes HADES operation and maintenance.

## 1 Introduction

The frontiers of physical exploration have come down to understanding the infinitely small (quantum mechanics) and the infinitely large (astrophysics): the physics of extremes. Everything in between has now virtually been covered. The next step in ground-breaking discoveries faces a big challenge: studying matter that does not exist naturally on Earth because it is too dense to be stable at atmospheric pressures or much too large to fit in a laboratory. In the hope to solve this problem, scientists need to develop new instruments able to handle an even higher degree of energy density required to produce such matter. From discovering the macroscopic properties of strongly interacting quantum systems to constraining the physics of supersonic astrophysical jet, scientists are faced with fundamental questions that can only be answered by controlled experiments done in the laboratory.

Historically, kilojoule lasers[1,2,3] or heavy ion beams[4] have been used to tackle these problems. Pulsed-power drivers have been surprisingly absent from this scientific race. However, they are perfectly fit to impact this research frontier. On the one hand, high power lasers excel at producing extremely intense, short-lived bursts of energy, which strongly interact with electrons. On the other hand, pulsed-power drivers produce gradual power ramps (<200ns), ideally suited to interact with bulk matter, allowing measurements of macroscopic material properties like viscosity or heat conduction. Furthermore, unlike lasers and accelerators which rely on inertial confinement, the magnetic field generated by pulsed-power drivers confines dense and hot matter on longer time scales (>10ns). Pulsed-power systems are also more cost effective than lasers or accelerators. Yet the true advantage of pulsed-power drivers has become apparent with the arrival of linear transformer drivers[5] (LTD). LTDs allow pulsed-power drivers to be transportable. Pulsed-power drivers can be relocated next to best possible x-ray sources, from synchrotron to X-ray free-electron lasers (XFEL) to probe the densest possible states of matter.

To this end, we are developing a 250 GW pulsed-power driver, called HADES (High Amperage Driver for Extreme States), capable of producing large volumes of matter under extreme conditions, confining this matter long enough to study its equilibrium properties. The design of HADES follows three simple design rules to fully take advantage of light sources such as LCLS[6]:

- The geometry has to be open, side-on and end-on diagnostic access are necessary;
- The driver can be transported and integrated into large x-ray source facilities;
- The repetition rate has to be close to one shot every ½ hour.

Three unique characteristics set this driver apart from other pulsed-power generators. First, the LTD technology allow HADES to be extremely compact, yet it can produce 1 MA of current in 150 ns into a load with an inductance of 20nH. Second, HADES is portable. Third, HADES is modular. It can be split into two drivers that can be used at two different locations. After this introduction, we present HADES main characteristics, then the electrical design is explained. We also highlight the mechanical properties of a single LTD. Operation and maintenance are also described.

# 2 System design

## 2.1 *Overall system design and capabilities*

The main idea behind HADES' design is compactness, an important factor in material procurement, ease of construction and facilitate maintenance. With compactness comes portability, allowing relocation of HADES to light source facilities and use x-ray probes to understand the properties of matter under extreme conditions. Table 1 gives HADES main parameters. HADES can deliver up to 1 MA of current in less than 150 ns. The load inductance must stay below 20nH, giving a total power of 250 GW. HADES footprint is relatively modest, less than 5m$^2$. The height of the machine is 1.6m. The total weight is less than 10 tons. HADES section is shown in Figure 1.

HADES is composed of two modules, each housing three LTD cavities. Each cavity weighs 1.5 tons. The main limitation of the cavity is the maximum operating voltage of 200 kV, rather than its current, limited to 1.1MA. As a result, a 1MA machine requires several cavities in series to drive practical experimental loads, with inductances on the order of 20 nH, adding to the 20 nH of transmission lines. When several cavities are connected in series, their internal inductances and resistances add up while their internal capacitance decreases. This overall trend rapidly reduces the available current that the cavity can drive, requiring to add cavities in parallel to regain the loss in current. The system becomes bulky very rapidly due to the adding of supplementary transmission lines.

To overcome this problem, we developed a set of nested transmission lines that adds the current of both modules near the load region using a post-hole (PH) convolute[7]. The top and bottom modules rest on top of each other. As shown in Figure 1, each module has its own anode and share a common cathode. The space between each anode and the cathode is adjusted so the top and bottom modules see the same impedance. Smaller inductive loads can be driven using similar configurations but require shorter transmission lines. As Table 2 shows, it is possible to run HADES with 1, 2 or 3 cavities per module. This modularity can prove useful if HADES need to be split between two facilities. These two configurations would require two separate sets of transmission lines, another vacuum vessel, charging supplies and trigger generators.

The design also aimed at reducing the impact of pulsed-power research on the environment. The spark gap switches use synthetic (i.e. dry) air instead of $SF_6$, a strong green-house gas. We also increased clearances between high voltage components to use silicone oil[8] instead of petroleum oil.

## 2.2 Electrical design

HADES' brick design, shown in Figure 2-a, was modeled using the accepted standard LRC model of a LTD[5,9]. Figure 3 shows the resistance, capacitance and inductance for such a brick using the design values provided by the different vendors[10,11,12,13]. The 22 bricks are assembled into the cavity presented in Figure 2-b. In turn this cavity is one of the six, shown in HADES' cross-section of Figure 1. Losses can be strongly reduced when using magnetic cores wound with nanocrystalline magnetic tape[14] rather that regular Metglass material. Each core is 0.125" thick. The cores are stacked on top of each other inside the cavity. The spark gap switches need to have very low inductance (L~50 nH). The casing is all stainless steel. The central insulator is made of rexolite and the outer insulator is using ultra-high molecular weight polyethylene. Each capacitor is connected to the cavity inner ring using brass strip line.

The electrical circuit model presented in Figure 3 is used to model a full cavity, where all 22 bricks are replaced by a set of equivalent components. Basically, all values were divided by 22 except for the equivalent capacitance which was multiplied by 22. The magnetic core values are not physically split in each brick, as shown in Figure 3. However, we used this representation to give Figure 3 a reasonable level of self-consistency. The inductances of the transmission lines connecting the two modules to the load were computed numerically from technical drawings using the formula

$$\frac{1}{2}LI^2 = \int_V \frac{B^2}{2\mu_0} dV. \tag{1}$$

If we ignore the post-hole convolutes, this integral can be computed easily when the system is axi-symmetric. The inductance of the post hole convolute is computed separately and added to the model next to the load. Figure 4 shows a simplified circuit model of HADES. While HADES can be fully modeled with four meta-components (i.e. one switch, one resistor, one inductor and one capacitor), we opted to keep the important topological features of the device visible in the model.

Figure 5 shows the time evolution of the current inside the load, together with the voltage at the load. No dynamic effects are considered in this simulation. In particular, the load inductance, the resistance and inductance of the spark gaps switches are constant. With the given parameters, the current rises to 1MA in 150 ns. The current rise stops after 200ns, reaching 1.02 MA. The maximum value of the current is limited by the load resistor, not by the load inductance. While real plasma loads are inductive, they are hardly resistive. Adding a physical resistor to the load is required in practice. This resistor damps the ringing of the LC circuit, limiting voltage and current reversal on the LTD capacitors. Depending on the experiment, the value of the resistor can be changed to match the load to the driver and obtain 1 MA of current.

The strip line connecting the capacitors to the inner rings are made with brass. Brass was used over copper for its higher rigidity and over stainless steel for its higher conductivity. As shown in Figure 2-b, each strip line is made from one single sheet of brass and bolted to the inner rings. The bolts in both strip lines are facing each other. The rexolite is in between. The screws need to be countersunk and covered with solder to eliminate electrical hot spots.

All electrical connections will be complemented with electrically conductive o-rings. We have tested different materials. Sacrificial o-rings made with tin solder has perform extremely well. The ductility of the materials allows for a tight electrical joint with no perceivable electrical hot spots.

## *2.3 Mechanical design*

The mechanical strength of the device and its portability requires a stiff casing. The whole casing is in stainless steel. The shape was kept as simple as possible to reduce the number of welds (which add unnecessary stresses to the system). The casing is made of three rings and two lids. The inner rings are 2" thick and will form the central pillar holding HADES together, as shown in Figure 1. The central insulators, located in between the inner rings, will carry the load down to the last inner which will rest on the ground. Rexolite was chosen since its compressive strength is sufficient to support the whole machine. However, this insulator requires to be locked in place between both rings to ensure mechanical stability. We designed the casing in such a way that the rexolite insulator is compressed by the top and bottom inner rings, rather than bolted into the rings. This design avoids the presence of hole in the insulator and the rings, as both locations are strongly stressed mechanically and electrically. Instead we used a clamping method where the top and bottom lids are pinned to the inner rings and the outer ring. It is possible to generate a clamping force when the height of the outer ring is shorter than the height of both inner rings and the central insulator. This force can be computed analytically[15]. Removing 6 mm to the total outer ring height generates one ton on the rexolite insulator. Each cavity is connected to the next using a stainless-steel ring which will hold two vacuum o-rings and two electrical o-rings. Similar rings are also used to connect the vacuum vessel and the lower vacuum port to the inner rings.

The vacuum vessel has two types of ports. (1) Large square windows (highlighted in yellow in Figure 6) are used for wide angle visible diagnostics such as laser shearing interferometry[16] and Faraday rotation to measure magnetic fields using polarimetry array or a laser wavefront analyzer[17,18,19]. Each laser beam can be split three ways allowing to get three time frames with one single laser. These diagnostics can allow us to measure plasma electron densities on the order of $10^{20}$ cm$^{-3}$ and magnetic field strengths higher than 0.5 T[20]. They also are used to gather Thomson scattered light[21] from a high power laser (10J, 1 ns) to measure plasma temperature and velocity. Square windows can be used to connect one or two independent X-pinch[22,23] pulsers to the vacuum vessel giving us x-ray backlighting capabilities to measure plasma densities and temperatures. (2) Round ports can be used for diagnostics with a narrower field of view like XUV framing cameras[24] (able to capture the dynamics of the hottest part of the plasmas, as well as XUV and x-ray spectrometers to measure electron density, temperature and magnetic field (via Zeeman splitting)[25].

HADES's shot rate is limited by the pump down time to reach a vacuum of $10^{-5}$ torr. To reach the desired repetition rate two large turbo-molecular pumps are required. One pump is located at the top of the vacuum chamber and one to pump at the bottom. Cryogenic pumps were not considered since the bulky apparatus they require to function will be a hindrance when operating and relocating HADES.

## 3 Operations

### 3.1 Charging supplies

The charging supply have been built recently and are capable to charge the capacitor banks up to +/-100 kV at 10 mA. The charge time for one cavity is on the order of 10 s for safe operations. It is not necessary to charge the system faster, but it is possible. The power supply uses a 20 kV/10 kHz transformer with a ferrite core. It is connected to a voltage multiplier, giving a dual output voltage of +/- 100 kV. The intrinsic current limit of the transformer is 10 mA. This limit was set to protect the voltage multiplier components which current limit is 200 mA. Technically, this system does not require charging resistors since the current is limited by design. However, all the capacitors inside one cavity are charged in parallel, the failure of one capacitor would be disastrous if charging resistors were not used. These resistors can also discharge the capacitors if HADES is not fired. These resistors should have a resistance on the order of 10MOhms which corresponds to 10 mA discharge current. To simplify design and augment reliability, we used conductive plastic instead of standard ceramic resistors. We tested different materials. Tivar Cleanstat performed extremely well. No resistor failure was found after 100,000 charges. Each resistor has a square cross-section of 1cm and a length of 30 cm. The disadvantage of conductive plastic compare to standard resistors is the resistivity of the material which can vary widely from batch to batch. However, we are interested in only high resistance, not its absolute value. Since all capacitors are charged from a single supply with virtually no current loss, a variation of the charging resistance will not affect the final charge of the capacitor.

Premagnetizing cores allows to reduce the amount of magnetic material required in the cavity. The pre-magnetizing charging supply should provide 50A DC. Each power supply which will be connected to each cavity core sets. One turn per set is sufficient to pre-magnetize each core.

## *3.2 Firing sequence*

Figure 7 shows the sequence of operations to run HADES. Time is given in minutes. All systems can be controlled by one single computer and one single master switch. The operation uses a computer for system check and the pre-shot sequence. The firing sequence is handled by this computer. The operator need to use a master (dead-man) switch to interrupt the sequence at any time, putting HADES into safe mode in case of system failure.

As HADES is fired one may expect some level of electromagnetic noise. A pulsed-power system is often seen as a source electromagnetic noise disturbing neighboring electronic circuits. However, LTD uses a completely hermetic stainless-steel casing, de facto acting as a Faraday cage. An electromagnetic pulse can only escape from the anode-cathode gap of the LTD itself, which opens directly onto the vacuum transmission line of HADES. Since both transmission lines are co-axial, the only location where the electromagnetic pulse can escape is near the load, which itself is surrounded by the vacuum vessel. In practice, this noise is minimal. For instance, the level of noise recorded a foot from MAIZE[26] shown in Figure 8 is on the order of 1mV into a 50Ohm impedance. The noise was measured using a Bdot probe with an area a couple of inches in diameter. This level of noise is easily shielded using standard laboratory practices, such as eliminating ground loops and recording electrical signals using coaxial cables. Electronic systems in close proximity of HADES may require minimal shielding.

## *3.3 Maintenance*

As mentioned earlier, the only two maintenance requirements are switches and capacitors. Both are rated for 50,000 shots. However, switches need regular cleaning (~10,000 shots) to prevent pre-fires. The shot rate indicated in Figure 7 means that we will have 3,600 shots per year. Switches will require cleaning after 2.5 years. Rather than dismantling the whole machine and working on all LTDs at once, we will spread the work by refurbishing one LTD every 6 months. For instance, if we are in the configuration LTD1-LTD2-LTD3-LTD4-LTD5-LTD6 before refurbishment (i.e. as shown in Figure 1), then the new sequence will be LTD2-LTD3-LTD4-LTD5-LTD6-LTD7 after refurbishment, where LTD7 is a spare LTD. The spare LTD has a dual purpose. First, it will be used to reduce HADES downtime so operations do not have to wait for the refurbished LTD to be put back inside the module. Second, if we have a major LTD failure (e.g. internal capacitor short) during a scientific campaign, the dead LTD will be pulled out of HADES and will be replaced with the spare LTD. The spare LTD will be fully tested and will be fired several times (~100) to condition the switches.

## 3.4 Relocation

HADES portability is a unique asset that will allow us to best diagnose ultra-dense matter using existing facilities such as LCLS. We reduced the size of the 1MA driver so it can be moved relatively easily. If the hosting facility has an overhead crane or a forklift capable of lifting 1.5 ton, HADES can be assembled in less than one month. A cart has been specifically designed to move LTDs vertically through standard building doors (shown in Figure 9).

# 4 Conclusion

This paper discussed a new pulsed-power driver called HADES. This pulsed-power driver, based on LTD technology is compact, simple to operate and can be relocated to different facilities depending on research campaigns and their diagnostic needs. The system uses a novel transmission-line geometry, where LTDs can be connected in series or in parallel, with very little impact on the driver footprint. This design gives unprecedented flexibility since:

- The geometry is open, allowing for side-on and end-on diagnostic access;
- The driver can be transported and integrated into large x-ray source facilities;
- The repetition rate is less than ½ hour, enabling a fast turnaround time.

While HADES can generate 1MA of current inside a 20 nH load, each LTD is only 2m in diameter, greatly simplifying the construction, maintenance and operation of the driver compared to conventional like Marx-bank based generators. While such systems are more flexible with respect to load geometry, their operation and maintenance are problematic. If the science can accommodate the inductive constraints imposed to the load, an LTD-based system becomes a practical and attractive tool to study matter under extreme conditions. Not only maintenance and operations are greatly simplified, but relocation and integration to other diagnostic facilities become possible. Even if the driver's ultimate purpose is to stay at a single location, the requirements on the laboratory, such as door size, power supply, floor load or ceiling height are relatively minor, allowing many institutions to add a 250 GW system to their research portfolio with very little change in their infrastructure.

**Acknowledgements**: This research was supported by the DOE grant number DE-SC0016252.

| HADES | Value |
|---|---|
| Nominal current rise time (ns) | 150 |
| Total current (MA) | 1 |
| Maximum load inductance (nH) | 20 |
| Cavity diameter (m) | 2.1 |
| Power (GW) | 250 |
| Total Energy (kJ) | 75 |
| Total machine mass (tons) | 10 |
| Total machine height (m) | 1.6 |

Table 1. Hades parameters

| Number of LTDs per module | Maximum load parameters | |
|---|---|---|
| 3 | $R_{load}(m\Omega)$ | 200 |
| | $L_{load}(nH)$ | 20 |
| 2 | $R_{load}(m\Omega)$ | 140 |
| | $L_{load}(nH)$ | 14 |
| 1 | $R_{load}(m\Omega)$ | 70 |
| | $L_{load}(nH)$ | 7 |

Table 2. Modular HADES. All configurations generate 1MA with a current rise time of 200 ns

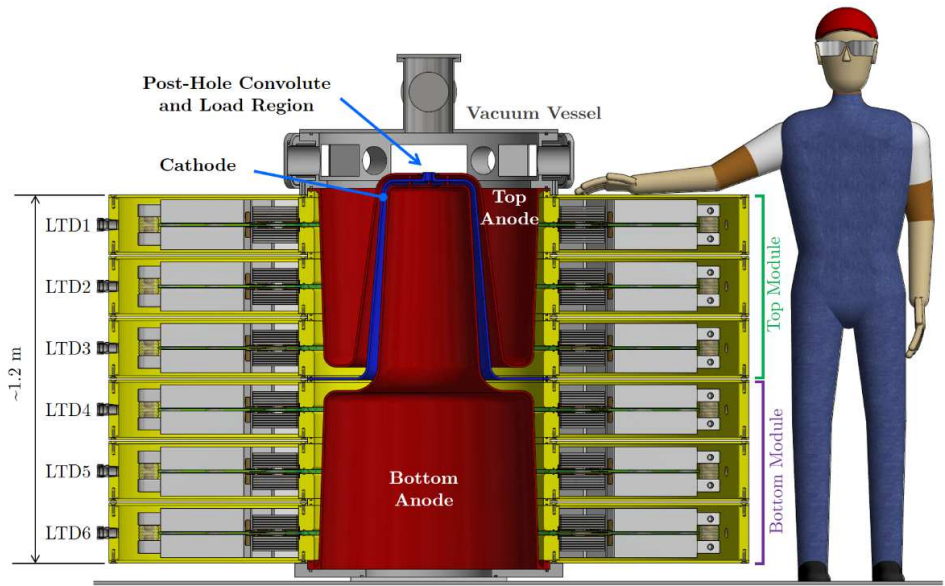

Figure 1. A cross-section of HADES. Both anodes are red. The cathode is blue. The vacuum vessel and the pumping manifold are gray. No gas line, bolt, O-ring, gas line, structural insulator, electrical wire is shown.

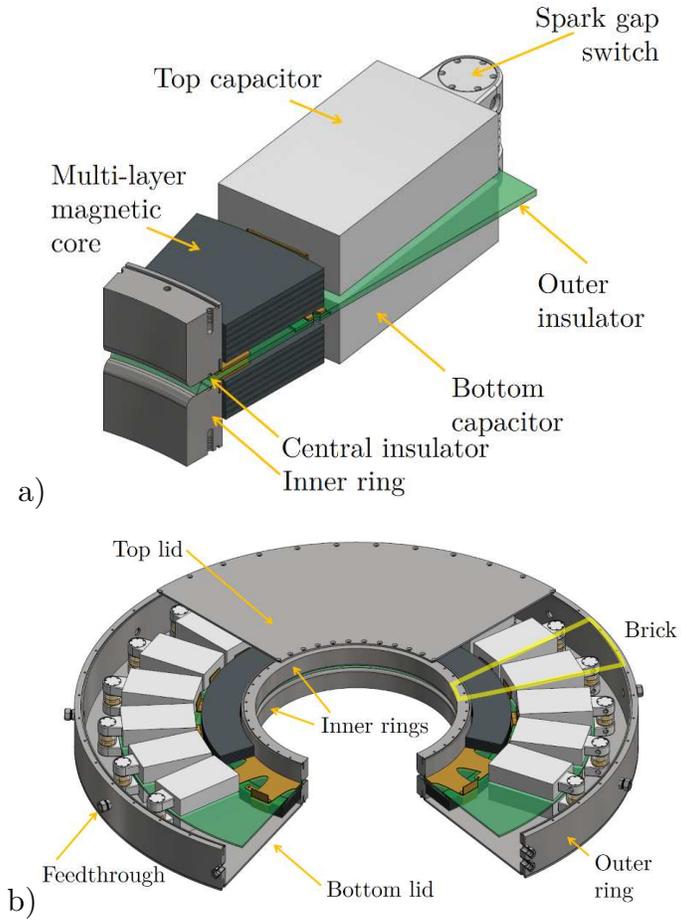

Figure 2. a) HADES brick design with two 80 nF capacitors and one spark gap switch. b) One of HADES LTDs. The LTD has been cut to show its internal components. One brick is highlighted by a yellow pie slice. No bolt, O-ring, gas line, electrical wire and insulator is shown.

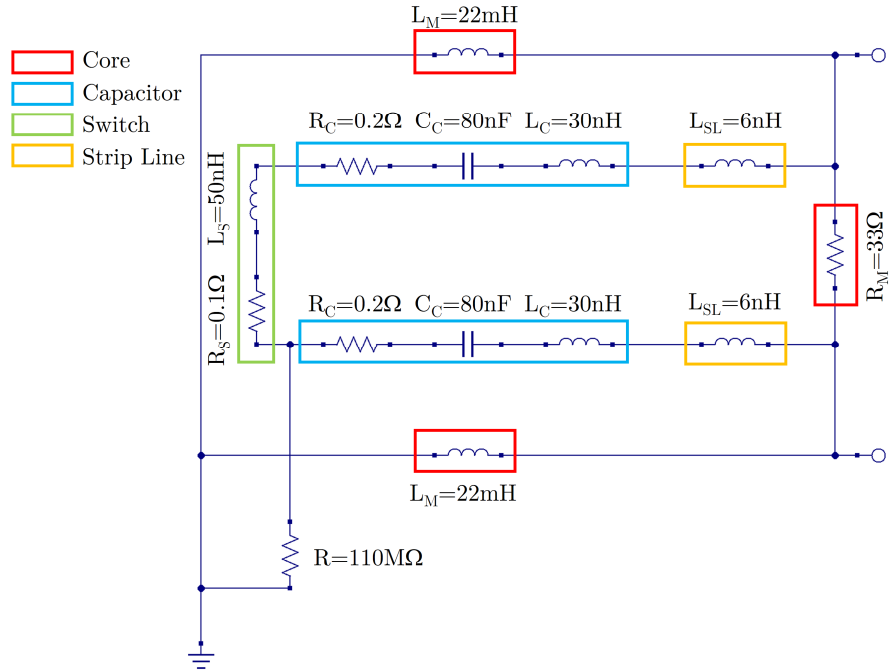

Figure 3. Brick electrical model used to model one cavity. The core inductance and loss resistance inside the brick model are 22 times larger than the total inductance and loss resistance of the core. Each cavity is made of 22 brick models in parallel. The large resistor connected to ground is used to stabilize SPICE's numerics.

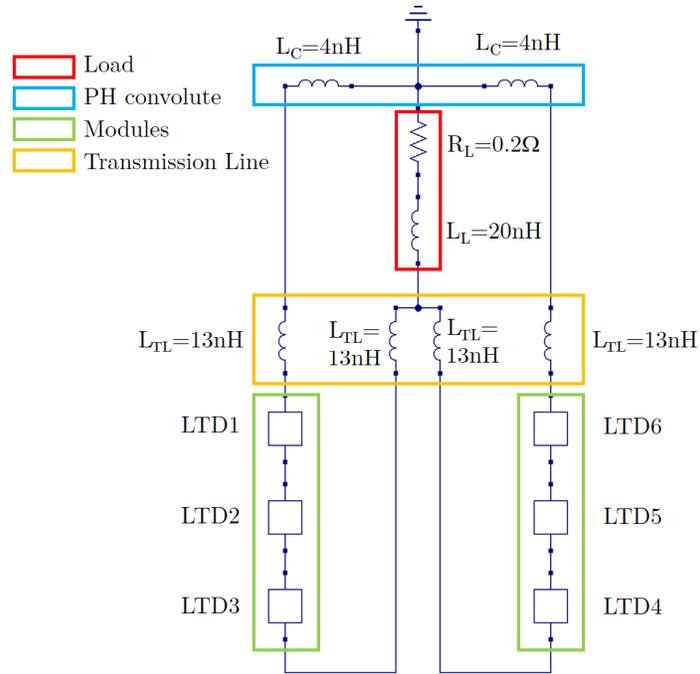

Figure 4. HADES electrical model, including the post hole (PH) convolute.

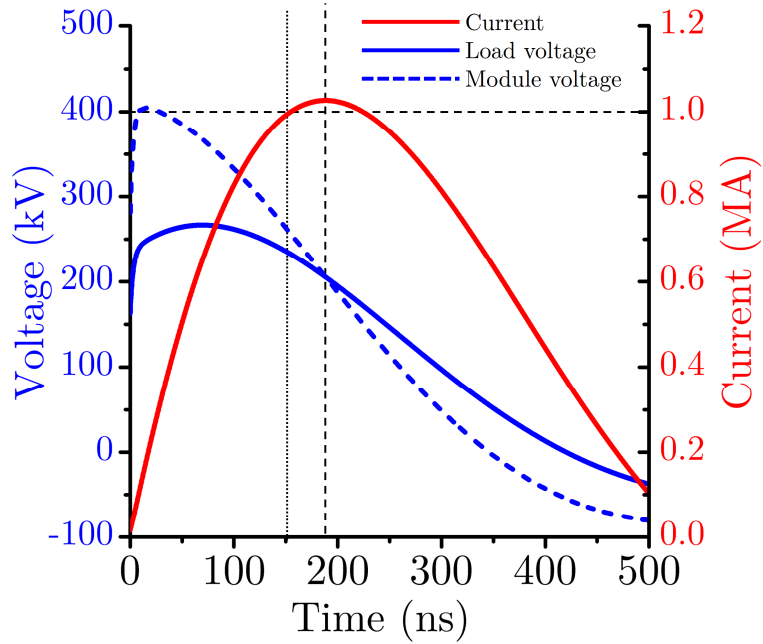

Figure 5. Spice simulation of HADES showing the time evolution of the current (red) and voltage (blue) for a load with a resistance of 0.2 ohm and an inductance of 20 nH. The dashed lines indicate the nominal current and rise time. The vertical dotted line indicates the time at which HADES reaches 1 MA.

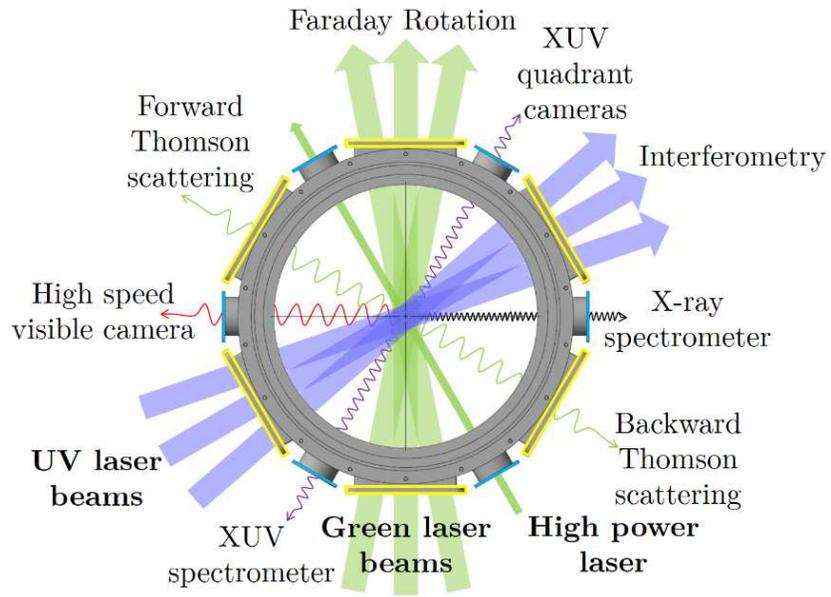

Figure 6. Diagnostics distribution around a top view of the vacuum vessel. The square windows are highlighted in yellow and the round port in light blue.

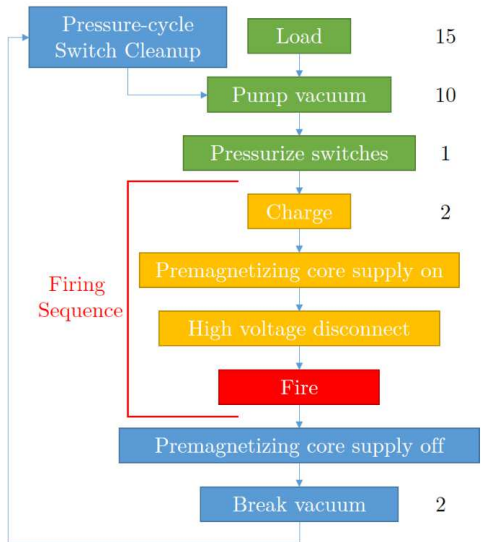

Figure 7. Operation of HADES with duration (min.) showing the pre-shot (green) and post-shot (blue) sequence. The main shot sequence (orange/red) duration is constrained only by capacitor charge time.

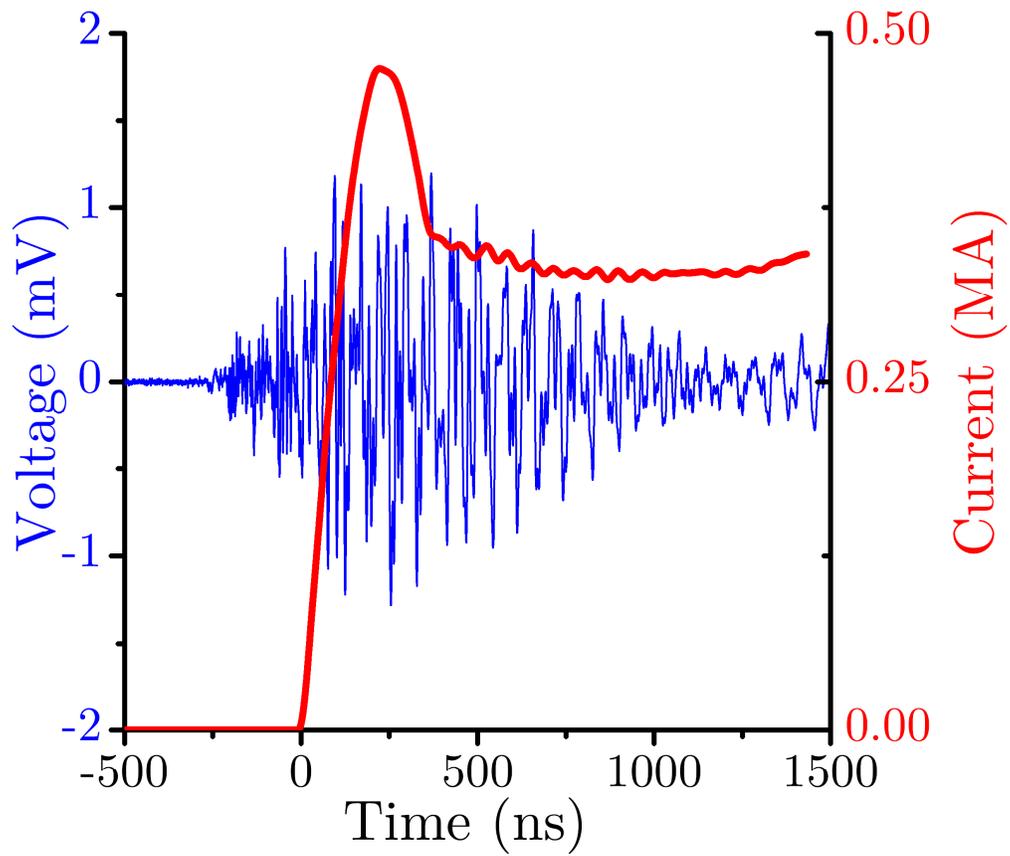

Figure 8. Electromagnetic pulsed recorded by a Bdot loop next to MAIZE connected into a 50Ohm. The loop had an approximate area of 4 square inches.

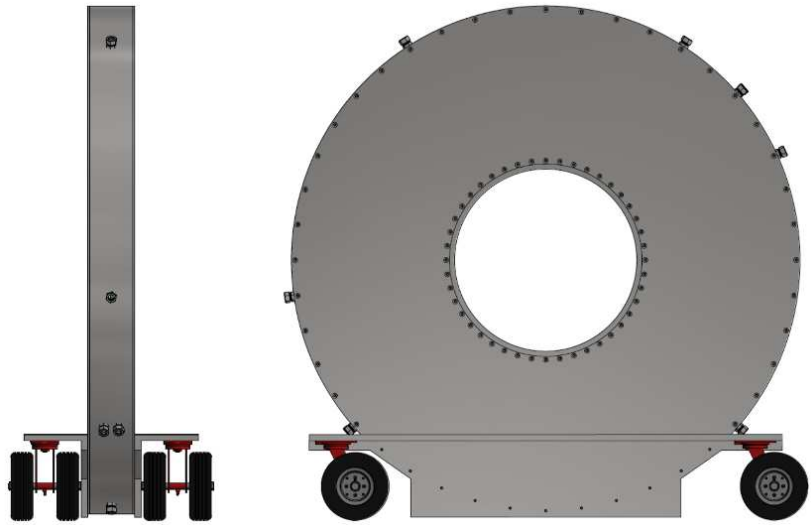

Figure 9. Front and side view of HADES low profile cart. The left and right wheel assemblies are bolted directly onto the LTD lid using bolt holes inside the LTD casing. The wheels can be inflated or deflated to allow the LTD to go through doors of slightly different heights. This cart will be used to store LTDs in the XSPL.